\documentclass[superscriptaddress,preprintnumbers,amsmath,amssymb,nofootinbib,dvipdfmx,eqsecnum,a4paper,secnumarabic,11pt]{revtex4}         
\usepackage{graphicx}
\usepackage[dvips]{color}
\usepackage{latexsym}    
\usepackage{mathrsfs}    
\usepackage{hyperref}    
\usepackage{titlesec}    
\newcommand{\tr}{\mbox{tr}}
\newcommand{\VEV}[1]{\langle #1 \rangle}

\newcommand*{\justifyheading}{\raggedright}
\titleformat{\chapter}[display]
  {\normalfont\huge\bfseries\justifyheading}{\chaptertitlename\ \thechapter}
  {20pt}{\Huge}
\titleformat{\section}
  {\normalfont\Large\bfseries\justifyheading}{\thesection}{1em}{}
\titleformat{\subsection}
  {\normalfont\large\bfseries\justifyheading}{\thesubsection}{1em}{}
\titleformat{\subsubsection}
  {\normalfont\bfseries\justifyheading}{\thesubsubsection}{1em}{}
\usepackage[english]{babel}

\begin{document}

%----------------------------------------------------------------------
% Title
%----------------------------------------------------------------------
\preprint{}
\preprint{}

\title{
Emergent two-Higgs doublet models
}

\author{Tomohiro Abe}
\affiliation{
Institute for Advanced Research, Nagoya University,
Furo-cho Chikusa-ku, Nagoya, Aichi, 464-8602 Japan
}
\affiliation{
Kobayashi-Maskawa Institute for the Origin of Particles and the
Universe, Nagoya University,
Furo-cho Chikusa-ku, Nagoya, Aichi, 464-8602 Japan
}
\author{Yuji Omura}
\affiliation{
Kobayashi-Maskawa Institute for the Origin of Particles and the
Universe, Nagoya University,
Furo-cho Chikusa-ku, Nagoya, Aichi, 464-8602 Japan
}

\begin{abstract}
We investigate origin of three features that are often assumed in analysis of
 two-Higgs doublet models:
(i) softly broken $Z_2$ symmetry, 
(ii) CP invariant Higgs potential, 
and (iii) degenerated mass spectra.
We extend electroweak gauge symmetry, introducing extra gauge symmetry and
extra scalars, and 
we show that our models effectively derive two-Higgs doublet models at 
low energy which naturally hold the three features.
We also find that the models can solve the strong CP problem.

\end{abstract}

\maketitle

\section{Introduction}

The Standard Model (SM) is elaborately constructed and  
succeeds in reproducing almost all experimental results so far.
The model is based on a SU(3)$_c \times$ SU(2)$_L \times$ U(1)$_Y$ gauge theory, 
and the unification of the weak and electromagnetic interactions is achieved by
introducing a Higgs doublet and spontaneous electroweak (EW) symmetry breaking. 
 In 2012, the signal predicted by the SM Higgs was finally found at the LHC \cite{Aad:2012tfa,Chatrchyan:2012ufa}, so that we are sure that the Higgs particle really exists in our nature. 

On the other hand, we expect that the SM is an effective model and new physics behind the SM is also discovered near future.
One reason is that the SM has several non-trivial structures. For instance, the anomaly-free conditions for the gauge symmetries
are achieved by the very non-trivial charge assignment of the quarks and leptons.
Besides, the Higgs couplings with the fermions are unnaturally hierarchical and we do not know 
the origin of the Higgs potential to trigger the EW symmetry breaking.
We note that some deviations from the SM predictions in some experiments have been reported,
so they also motivate new physics beyond the SM.

 In fact, many extensions of the SM have been proposed so far. However, it 
 is not so easy to consider the extensions, because of the stringent experimental constraints
 and the non-trivial structures of the SM. As we mentioned above, the anomaly-free conditions
 are satisfied miraculously, so that it is not simple to introduce new fermions charged under the SM gauge groups.
 Furthermore, the experimental constraints are getting stronger according to the precision measurements.
Especially, the SM is very successful in flavor physics and we know 
that new physics should avoid tree-level Flavor Changing Neutral Currents (FCNCs).  

Two-Higgs doublet model (2HDM) is one of the simple extended models without such serious problems.
In this extension, one extra Higgs doublet is introduced, and the anomaly-free conditions are the same as the SM ones.
In fact, 2HDMs have been widely discussed so far motivated by the experimental anomalies, such as the muon $g-2$~\cite{hep-ph/0102297, hep-ph/0009292, hep-ph/0103183,%hep-ph/0103223,
hep-ph/0302111, hep-ph/0104056, 0808.2509, 0909.5148, 
1409.3199, 1412.4874, 1502.04199, 1504.07059, 1511.05162, 1605.06298}.
Besides, many Beyond Standard Models (BSMs) to solve the theoretical issues in the SM predict such an extra Higgs doublet, so that 2HDMs have been well analyzed as the low-energy effective models, although it is often difficult to derive 2HDMs effectively. This is because 
extra other matters generally reside at the same scale as the extra Higgs doublet in many BSMs.

Even in such a simple BSM, however, some assumptions are usually supposed 
to avoid the conflicts with the experimental results. For instance, there are generally tree-level FCNCs
in 2HDMs, if the two Higgs doublets are not distinguishable. 
We usually assign a softly broken $Z_2$ symmetry to avoid the tree-level FCNCs,
and then we only allow the minimal flavor violation (MFV)~\cite{Glashow:1976nt}.
%
%
%
% Moreover, there is a possibility that the Higgs potential is CP violating,
% so that it is often assumed to be CP invariant to avoid additional CP violating source 
% in the lagrangian. 
%
Moreover, the Higgs potential contains CP violating terms in
general. This is an attractive point of 2HDM, for example, for
electroweak baryogenesis scenario. Nevertheless, 
CP invariance is often assumed in 2HDM.
In addition, the results in the precision measurements of the EW interaction should be
consistent with the predictions of 2HDMs. 
For example, the EW precision tests suggest that $\rho$ parameter is close to one.
%This requires unnaturally small dimensionless couplings, namely
%$\lambda_4$ and $\lambda_5$, in 2HDMs. 
This requires unnaturally degenerated dimensionless couplings, namely
$\lambda_4 = \lambda_5$, in 2HDMs.

These three assumptions are required by the experimental results.
If there is an extra Higgs doublet in our nature, these three features
would offer an important clue as to the unknown physics behind the SM.
In this paper, we investigate a possible underlying theory which effectively derives the 2HDM
respecting the three features  at low energy. 
Especially, we study some models where
the electroweak gauge symmetry is extended\footnote{Another possibility
is to impose some symmetries on the Higgs potential~\cite{Dev:2014yca}.}, and show that the models
naturally have the three features often assumed. 
We note that 2HDMs with gauged U(1) symmetry have been proposed
as the origin of the softly broken $Z_2$ \cite{Ko-2HDM,Ko-2HDM1,Ko-2HDM2}.
This U(1) symmetry may be originated from the grand unified theory \cite{Ko-2HDM2}, but the setup is complicated and the $\rho$ parameter is deviated from one at the tree level  \cite{Ko-2HDM,Ko-2HDM1,Ko-2HDM2}.
Compared to this proposal, our models in this paper are much simple
and can avoid too large deviation of the $\rho$ parameter.

The rest of this paper is organized as follows.
In section~\ref{sec:2hdm}, we briefly review the 2HDM.
In section ~\ref{sec:model}, we propose a model with extended electroweak gauge symmetry. 
In section~\ref{sec:type-1_2HDM}, we derive the low energy effective theory
of the model discussed in Sec.~\ref{sec:model}, and show that it behaves
as the type-I 2HDM with softly broken $Z_2$ symmetry without any CP phases in
the Higgs potential. The model also predicts degenerated mass spectra for
the CP-odd and the charged Higgs bosons. Thus all the three conditions are automatically satisfied.
In section~\ref{sec:type-3_2HDM}, we extend the model to lead 2HDM
other than the type-I 2HDM.
Section~\ref{sec:summary} is devoted to our conclusion.

\section{Review of two-Higgs doublet models}
\label{sec:2hdm}

In this section, we review the two-Higgs doublet model with the softly
broken $Z_2$ symmetry widely discussed.
We have two Higgs fields, $\Phi_1$ and $\Phi_2$, charged under SU(2)$_L \times$U(1)$_Y$.
In general the Higgs potential at the renormalizable level is given as follows:
\begin{align}
 V=&
m_1^2 \Phi_1^{\dagger} \Phi_1
+
m_2^2 \Phi_2^{\dagger} \Phi_2
-
\left(
m_3^2
\Phi_1^{\dagger} \Phi_2
+
(h.c.)
\right) \nonumber 
\\
&
+ \frac{1}{2} \lambda_1 (\Phi_1^{\dagger} \Phi_1)^2
+ \frac{1}{2} \lambda_2 (\Phi_2^{\dagger} \Phi_2)^2
+ \lambda_3 (\Phi_1^{\dagger} \Phi_1) (\Phi_2^{\dagger} \Phi_2)
+ \lambda_4 (\Phi_1^{\dagger} \Phi_2) (\Phi_2^{\dagger} \Phi_1) \nonumber 
\\
&
+
\left(
\frac{1}{2}
\lambda_5
 (\Phi_1^{\dagger} \Phi_2)^2 
+
\lambda_6
 (\Phi_1^{\dagger} \Phi_1)  (\Phi_1^{\dagger} \Phi_2)
+
\lambda_7
 (\Phi_2^{\dagger} \Phi_2)  (\Phi_1^{\dagger} \Phi_2)
+
(h.c.)
\right)
.
\label{eq:2hdm_potential}
\end{align}
Four parameters, $m_3^2$, $\lambda_{5}$, $\lambda_{6}$, and
$\lambda_{7}$, can be complex, and they are CP violating. 
Now, we impose a softly broken $Z_2$
symmetry to the Higgs fields: $(\Phi_1, \Phi_2) \to (\Phi_1, - \Phi_2)$.
The $Z_2$ symmetry forbids $\lambda_6$ and $\lambda_7$ terms.
The $m_3^2$ term breaks the $Z_2$ symmetry softly, but
can shift the scalar masses.
% When they are treated as complex
% numbers, some people call the model as ``complex 2HDM''[ref].
Let us define the vacuum expectation values (VEVs) of the Higgs fields as
\begin{align}
 \VEV{\Phi_1} = \frac{v_d}{\sqrt{2}},\ \
 \VEV{\Phi_2} = \frac{v_u}{\sqrt{2}},
\end{align}
and then the relation with the Fermi constant is
\begin{align}
v_u^2 + v_d^2
=
 \frac{1}{\sqrt{2} G_F} 
\equiv
v^2
\simeq
(246\text{~GeV})^2
.
\end{align}
We also define $\beta$ as follows:
\begin{align}
 \cos\beta = \frac{v_d}{v},\ 
 \sin\beta = \frac{v_u}{v}.
\end{align}

We have eight degrees of freedom in the scalar fields and three of them
are eaten by the gauge bosons, and thus we have five physical
states: two of them are CP-even states ($h$, $H^0$), one is a CP-odd state
($A^{0}$), and the others are a pair of charged scalar ($H^{\pm}$).
Their masses are given by the parameters in the Higgs potential. 
%In the following, the masses of the CP-odd and the charged Higgs are important.
When the $Z_2$ symmetry is imposed, the charged and CP-odd Higgs masses
satisfy 
\begin{align}
 m_A^2 =& M^2 - \lambda_5 v^2, \\
 m_{H^{\pm}}^2 =& m_A^2 + \frac{\lambda_5 - \lambda_4}{2} v^2,  \label{eq;massdifference}
\end{align}
where
\begin{align}
 M^2 =&  \frac{m_3^2}{\sin\beta \cos\beta}.
\end{align}
Here $\lambda_5$ is assumed to be real.

There are four different assignments of the $Z_2$ symmetry to
fermions~\cite{Barger:1989fj, hep-ph/9401311, 0902.4665}. The assignments and
the model names are summarized in Table~\ref{tab:Z2}. 
These assignments forbid FCNCs involving neutral scalars. The physics in the each 2HDM has been widely
studied, although their origins of $Z_2$ symmetries are unclear.
%
%
%======================================================================
% table for matter contents
%======================================================================
\begin{table}[t]
\centering
\caption{$Z_2$ assingment of the matters in 2HDM. If the model not
 assign the $Z_2$ symmetry is called type-III.}
\label{tab:Z2}
\begin{tabular}{c|cc|cc|ccc}
  & $\Phi_1$ & $\Phi_2$ & 
$q_L$ & $\ell_L$ & $u_R$ & $d_R$ & $\ell_R$ \\
\hline \hline
type-I & + & $-$ & + & + & $-$ & $-$ & $-$ \\
type-II & + & $-$ & + & + & $-$ & $+$ & $+$ \\
type-X & + & $-$ & + & + & $-$ & $-$ & $+$ \\
type-Y & + & $-$ & + & + & $-$ & $+$ & $-$ \\
\hline \hline
type-III & \multicolumn{7}{|c}{no $Z_2$ assingment}
\end{tabular}
% \\
% \begin{tabular}{c|cccc}
%  & tyep-I & type-II & type-X & type-Y \\
% \hline
% $\Phi_1$ &  +   & +  & + & +  \\
% $\Phi_2$ &   $-$   &  $-$  &  $-$ &  $-$ \\
% \hline
% $q_L$ &  +   & +  & + & +  \\
% $\ell_L$ &  +   & +  & + & +  \\
% $u_R$ &  $-$   &  $-$  &  $-$ &  $-$  \\
% \hline
% $d_R$ &   $-$  & +  &   $-$ & +  \\
% $\ell_R$ &   $-$  & +  &  + & $-$  
% \end{tabular}
\end{table}
%======================================================================

In 2HDMs, the constraint of the $\rho$ parameter, or $T$ parameter, is important.
It depends on the scalar masses and $WWh$-coupling ($g_{WWh} = \kappa_V g_{WWh}^{SM}$).
%Note that $g_{WWh}^{SM}$ is the SM one of the  $WWh$-coupling.
Since the LHC result implies that
the coupling is almost the same as the SM prediction, we assume $\kappa_V
\simeq 1$. Then the $T$ parameter is approximately evaluated as
\begin{align}
 \alpha T \simeq&
\frac{1}{16 \pi^2 v^2}
\Biggl(
m_{H^0}^2
+
\frac{m_A^4 (m_{H^{\pm}}^2 - m_{H^0}^2)}{(m_A^2 - m_{H^{\pm}}^2)(m_A^2 - m_{H^0}^2)}
\ln \frac{m_{H^{\pm}}^2}{m_A^2} 
-
\frac{m_{H^0}^4 (m_{H^{\pm}}^2 - m_{A}^2)}{(m_{H^0}^2 - m_{H^{\pm}}^2)(m_A^2 - m_{H^0}^2)}
\ln  \frac{m_{H^{\pm}}^2}{m_{H^0}^2} 
\Biggr)
.
\end{align}
We easily find that the right-hand side is vanishing when either $m_A = m_{H^{\pm}}$ or $m_{H^0} = m_{H^{\pm}}$ is satisfied~\cite{Gerard:2007kn}.
Since the right-hand side should be less than ${\cal O}(10^{-3})$ according to the electroweak precision measurements~\cite{Agashe:2014kda}, two of the masses should be highly degenerated. 
This degeneracy requires a parameter tuning in the Higgs potential. For example $\lambda_4
\simeq \lambda_5$ is required for $m_A \simeq m_{H^{\pm}}$. 
It is unclear whether a mechanism which makes $\lambda_4 \simeq \lambda_5$
exists or not within the 2HDM.

%%%%%%%%%%%%%%%%%%%%%%%%%%%%%%%%%%%%%%%%%%%%%%%%%%%%%%%%
\section{Emergence of two-Higgs doublet models satisfying the three conditions }
\label{sec:model1}
In this section, we consider the extension of the SM that leads
2HDMs at low energy. Our effective 2HDM will satisfy the following three conditions to evade the stringent
experimental constraints:
\begin{enumerate}
\renewcommand{\labelenumi}{(\roman{enumi})}
\item
 softly broken $Z_2$ symmetry is remained, 
\item
 Higgs potential is CP invariant, 
\item
Two scalars are degenerate.
\end{enumerate}

At first, we consider a model which effectively realizes the type-I 2HDM.
Note that only one Higgs doublet couples with all fermions and the other
Higgs doublet does not in the type-I 2HDM.
Depending on the vacuum alignment of the Higgs fields,
the model can also leads to the Inert 2HDM~\cite{Deshpande:1977rw,
Barbieri:2006dq} , where the lightest neutral scalar becomes stable and
a dark matter candidate.

\subsection{A model with extended electroweak gauge symmetry}
\label{sec:model}
%\ref{sec:model}
Now, we consider a BSM with extended gauge symmetry, based on Ref.~\cite{1305.2047}.
In this model, the electroweak gauge symmetry is extended to
SU(2)$_0 \times$SU(2)$_1 \times$U(1)$_2$. We introduce three scalar
fields, $H_1$, $H_2$ and $\phi_3$, for the gauge symmetry breaking. Their quantum numbers are
summarized in Table~\ref{tab:matters}.
$q^i_L$, $u^i_R$ and $d^i_R$ denote left-handed and right-handed quarks respectively.
 $\ell^i_L$ and $e^i_R$ are left-handed and right-handed leptons.
 The left-handed matter fields are charged under the SU(2)$_0$ gauge symmetry.
 
%
%
%
%======================================================================
% table for matter contents
%======================================================================
\begin{table}[tb]
\centering
\caption{Quantum numbers of the Higgs and matter fields ($i=$1, 2, 3).}
\label{tab:matters}
\begin{tabular}{c|ccc|cc|ccc}
& $H_1$ & $H_2$ & $\phi_3$ & 
$q^i_L$ & $\ell^i_L$ &
$u^i_R$ & $d^i_R$ & $e^i_R$ \\
\hline \hline
SU(2)$_0$ & $\bf 2$ & $\bf 1$ & $\bf 2$ &
$\bf 2$ & $\bf 2$ &
$\bf 1$ & $\bf 1$ & $\bf 1$ \\
SU(2)$_1$ & $\bf 2$ & $\bf 2$ & $\bf 1$ &
$\bf 1$ & $\bf 1$ &
$\bf 1$ & $\bf 1$ & $\bf 1$ \\
U(1)$_2$ & $0$ & $\frac{1}{2}$ & $\frac{1}{2}$ &
$\frac{1}{6}$ & $-\frac{1}{2}$ &
$\frac{2}{3}$ & $-\frac{1}{3}$ & $-1$ \\
SU(3)$_c$ & $\bf 1$ & $\bf 1$ & $\bf 1$ &
$\bf 3$ & $\bf 1$ &
$\bf 3$ & $\bf 3$ & $\bf 1$ 
\end{tabular}
% \begin{tabular}{c|cccc}
% fields & SU(2)$_0$ & SU(2)$_1$ & U(1)$_2$ & SU(3)$_c$\\
% \hline
% $H_1$ & $\bf 2$  & $\bf 2$ & 0 & 1  \\
% $H_2$ & $\bf 1$  & $\bf 2$ & 1/2 & $\bf 1$ \\
% $\phi_3$ & $\bf 2$  & $\bf 1$ & 1/2 & $\bf 1$  \\
% \hline
% $q_L$ & $\bf 2$ & $\bf 1$ & 1/6 & $\bf 3$ \\
% $\ell_L$ & $\bf 2$ & $\bf 1$ & -1/2 & $\bf 1$ \\ 
% \hline
% $u_R$ & $\bf 1$ & $\bf 1$ & 2/3 & $\bf 3$ \\
% $d_R$ & $\bf 1$ & $\bf 1$ & -1/3 & $\bf 3$ \\
% $e_R$ & $\bf 1$ & $\bf 1$ & -1 & $\bf 1$ \\
% \end{tabular}
\end{table}
The gauge symmetric Yukawa couplings, which generate the mass matrices for the matters, are given by
\begin{equation}
{\cal L}^{\text{Yukawa}}
= 
- 
\sum_{i,j}
y_u^{ij}
\overline{q_L^{i}} 
\widetilde{\phi_3}
 u_R^{j}
- 
\sum_{i,j}
y_d^{ij}
\overline{q_L^{i}} 
\phi_3
 d_R^{j}
- 
\sum_{i,j}
y_e^{ij}
\overline{\ell_L^{i}} 
\phi_3
 e_R^{j}
+
h.c..
\end{equation}

The scalar potential is written down at the renormalizable level: 
\begin{align}
 V(H_1, H_2, \phi_3)
=&
  \mu_1^2 \tr\left( H_1 H_1^{\dagger} \right)
+ \mu_2^2 H_2^{\dagger} H_2
+ \mu_3^2 \phi_3^{\dagger} \phi_3  \nonumber
\\
&
+
\frac{1}{2}
\left( 
 \kappa 
\phi_3^{\dagger} H_1 H_2
+
(h.c.)
\right)  \nonumber
\\
&
+ 
\widetilde{\lambda}_1 
\left( \tr\left( H_1 H_1^{\dagger} \right) \right)^2
+ 
\widetilde{\lambda}_2 
\left( H_2^{\dagger} H_2  \right)^2
+ 
\widetilde{\lambda}_3 
\left( \phi_3^{\dagger} \phi_3  \right)^2  \nonumber
\\
&
+ 
\widetilde{\lambda}_{12} 
\tr\left( H_1 H_1^{\dagger} \right)
\left( H_2^{\dagger} H_2  \right)
+ 
\widetilde{\lambda}_{23} 
\left( H_2^{\dagger} H_2  \right)
\left( \phi_3^{\dagger} \phi_3  \right)
+ 
\widetilde{\lambda}_{31} 
\left( \phi_3^{\dagger} \phi_3  \right)
\tr\left( H_1 H_1^{\dagger} \right) 
, \label{eq;HiggsPotential}
\end{align}
Note that $H_1$ is a two by two matrix charged under SU(2)$_0 \times$SU(2)$_1$ and is
defined as the field to satisfy 
\begin{align}
 \tau^2 H_1^* \tau^2& = H_1,
\end{align}
where $\tau^2$ is the second Pauli matrix.
The Higgs potential contains one complex parameter $\kappa$, but its
phase are eliminated by a field redefinition
of $H_2$. Therefore we can take all the parameters in the Higgs
potential as real numbers.
We can also take the VEVs of the Higgs fields as real thanks to the gauge
symmetries. Hence we do not have any source of the CP violation in the
Higgs sector.

When the heavy gauge bosons are extremely heavy, we can integrate them
out and construct a low-energy effective model. The effective model is 
similar to the two-Higgs doublet models with softly broken $Z_2$ symmetry. 
In the followings, we discuss the effective model
and see that it can be interpreted as the type-I 2HDM
or the Inert 2HDM satisfying our three conditions, (i)-(iii).

\subsection{Emergence of the type-I two-Higgs doublet model}
\label{sec:type-1_2HDM}
Here, we discuss our effective model predicted by the BSM with the extended gauge symmetry.
There are two SU(2) gauge symmetries, SU(2)$_0 \times$SU(2)$_1$, and one scalar, $H_1$, charged under
the both symmetries. Then, the VEV of $H_1$ breaks down the symmetries to ${\rm SU(2)}_L$:
$${\rm SU(2)}_0 \times {\rm SU(2)}_1 \to {\rm SU(2)}_L.$$
We parametrize $H_1$,
\begin{align}
 H_1 = \frac{1}{2} (v_1 + h_1) U_1,
\ 
\text{where}
\
U_1 = \exp\left( i \frac{\tau^a \pi_1^a}{v_1} \right). \label{eq;H1}
\end{align}
Here $v_1$ is the VEV of $H_1$. We choose appropriate parameters in the
Higgs potential to realize non-zero VEVs.
According to the symmetry breaking, Nambu-Goldstone (NG) bosons appear
and are eaten by the SU(2) symmetry orthogonal to SU(2)$_L$.
 
 We define the gauge fields for SU(2)$_0 \times$SU(2)$_1 \times$U(1)$_2$ as
 \begin{equation}
{\cal L}^{\text{gauge}}= 
- \frac{1}{4} \sum_{a=1}^{3} W^{a}_{0 \mu \nu} W^{a \mu \nu}_{0}
- \frac{1}{4} \sum_{a=1}^{3} W^{a}_{1 \mu \nu} W^{a \mu \nu}_{1}
- \frac{1}{4} B_{\mu \nu} B^{\mu \nu},
\end{equation} 
where the field strengths are depicted by the gauge fields, $W_{0 \, \mu}$, $W_{1 \, \mu}$ and $B_{ \mu}$ of SU(2)$_0 \times$SU(2)$_1 \times$U(1)$_2$:  $W_{0 \, \mu \nu}= \partial_\mu W_{0 \, \nu}- \partial_\nu W_{0 \, \mu}+ig_0  [W_{0 \, \mu}, \, W_{0 \, \nu}]$, and $W_{1 \, \mu \nu}=\partial_{\mu} W_{1 \, \nu} - \partial_{\nu} W_{1 \, \mu}
+i g_1 [W_{1 \, \mu}, \, W_{1 \, \nu}] $, and $B_{ \mu \nu}= \partial_{\mu} B_{ \nu} - \partial_{\nu} B_{\mu} $, respectively. 

The nonzero VEV of $H_1$ generates the mass term of the broken gauge symmetry,
which is a linear combination of $W_{0 \, \mu}$ and $W_{1 \, \mu}$,
and $\pi_1^a$ in $U_1$ are eaten by the gauge field.
 
For convenience, let us redefine the gauge field, $W_{1}^{\mu}$, as
\begin{align}
 V_{1}^{\mu}
=&
 U_1 W_{1}^{\mu} U_1^{\dagger} 
+
\frac{1}{i g_1} U_1 \partial_{\mu} U_1^{\dagger}
.
\end{align}
%
%  $V^\mu_1$ is transformed as  the SU(2)$_0$ triplet
% field with the gauge coupling of SU(2)$_1$, depicted by $g_1$. 
  $V^\mu_1$ is transformed by the SU(2)$_0$ gauge symmetry
 with the gauge coupling of SU(2)$_1$, depicted by $g_1$. 
It is useful to use $V^\mu_1$ instead
of $W^\mu_1$ in this
analysis. 
Then, we take the linear combinations of $W_{0}^{\mu}$ and $V_{1}^{\mu}$:
\begin{align}
 \rho_{\mu} &= c W_{0}^{\mu} - s V_{1}^{\mu}, \\
 W_{\mu} &= s W_{0}^{\mu} + c V_{1}^{\mu}, 
\end{align}
where the mixing is given by
\begin{align}
 c = \frac{g_0}{\sqrt{g_0^2 + g_1^2}}, \ \
 s = \frac{g_1}{\sqrt{g_0^2 + g_1^2}}.
\end{align}
In the effective model, the gauge symmetry is SU(2)$_L \times$ U(1)$_Y$,
and $W_\mu$ is the gauge boson of SU(2)$_L$, which is given by SU(2)$_0$
with a different gauge coupling.
Note that $\rho^{\mu}$ is a massive SU(2)$_L$ triplet.
Let us define the gauge couplings in the effective model:
\begin{align}
 g = \frac{g_0 g_1}{\sqrt{g_0^2 + g_1^2}}, \ \
 g_{\rho} = \sqrt{g_0^2 + g_1^2},
\end{align}
where $g$ is the SU(2)$_L$ gauge coupling associated with $W_{\mu}$.
Based on the above definitions, we find the correspondence between the
fields in the original and in the effective: 
\begin{align}
 W_{1}^{\mu \nu}
=&
 U_1 V_{1}^{\mu \nu} U_1^{\dagger},\\
%----------------------------------------
W_0^{\mu \nu}
=&
 s W^{\mu \nu} + c (D^{\mu} \rho^{\nu} - D^{\nu} \rho^{\mu})
+ i g_0 c^2 [\rho^{\mu}, \rho^{\nu}]
,\\
%----------------------------------------
V_1^{\mu \nu}
=&
 c W^{\mu \nu} - s (D^{\mu} \rho^{\nu} - D^{\nu} \rho^{\mu})
+ ig_1 s^2 [\rho^{\mu}, \rho^{\nu}]
,\\
%----------------------------------------
\hat{D}_{\mu} H_2
=&
U_1^{\dagger} (D_{\mu} \phi_2 - i g_1 s \rho_{\mu} \phi_2)
,\\
%----------------------------------------
\hat{D}_{\mu} \phi_3
=&
D_{\mu} \phi_3 + i g_0 c \rho_{\mu} \phi_3
,\\
%----------------------------------------
\hat{D}_{\mu} \psi_L
=&
D_{\mu} \psi_L + i g_0 c \rho_{\mu} \psi_L.
\end{align}
Here, $\phi_2$ is defined as
\begin{align}
 \phi_2 = U_1 H_2,
\end{align}
$\hat{D}_{\mu}$ is the covariant derivative with respect to SU(2)$_0
\times$SU(2)$_1 \times$U(1)$_2$,
and $D_\mu$ is the covariant derivative with respect to SU(2)$_L
\times$ U(1)$_2$.
%depending on the charge assignments of the fields. 
Namely, $D_\mu$ contains only
$W_{\mu}$ and $B_{\mu}$:
\begin{align}
 D_{\mu} \psi
=& (\partial_{\mu} + i g W_{\mu} + i g_2 Y B_{\mu}) \psi,\\
%------------------------------
 D_{\mu} \rho_{\nu}
=&
 \partial_{\mu} \rho_{\nu} + i g [W_{\mu}, \rho_{\nu}].
\end{align}
$\psi_{L}$ is the left-handed SM fermions charged under SU(2)$_0$ (SU(2)$_L$). 
$\psi_{R}$ is the right-handed.

Finally, we find our effective Lagrangian as follows,
\begin{align}
{\cal L}=& 
{\cal L}^{(0)} + {\cal L}^{(1)} + {\cal L}^{(2)} - V(h_1, \phi_2, \phi_3)
,
\label{eq:after_W'_gone}
\end{align}
where ${\cal L}^{(0)} $, ${\cal L}^{(1)} $, and ${\cal L}^{(2)} $ are given by
\begin{align}
{\cal L}^{(0)}
=&
+ \bar{\psi}_L i \gamma^{\mu} D_{\mu} \psi_L
+ \bar{\psi}_R i \gamma^{\mu} D_{\mu} \psi_R
- \bar{\psi}_L \tilde{\phi}_3  y_u \psi_R
- \bar{\psi}_L \phi_3  y_d \psi_R
+
(h.c.)
\nonumber\\
%------------------------------
&
+ \frac{1}{2} \partial_{\mu} h_1 \partial^{\mu} h_1
+ D_{\mu} \phi_2^{\dagger} D_{\mu} \phi_2
+ D_{\mu} \phi_3^{\dagger} D_{\mu} \phi_3
\nonumber\\
%------------------------------
&
- \frac{1}{2} \tr( W_{\mu \nu} W^{\mu \nu})
- \frac{1}{4} B_{\mu \nu} B^{\mu \nu}
,
\\
%==================================================
{\cal L}^{(1)}
=&
-g_0 c \bar{\psi}_L \gamma^{\mu} \rho_{\mu} \psi_L 
\nonumber\\
&
+ i g_1 s 
\left(
\phi_2^{\dagger} \rho_{\mu} D^{\mu} \phi_2
-
(D^{\mu} \phi_2^{\dagger}) \rho_{\mu} \phi_2
\right)
\nonumber\\
&
- i g_0 c 
\left(
\phi_3^{\dagger} \rho_{\mu} D^{\mu} \phi_3
-
(D^{\mu} \phi_3^{\dagger}) \rho_{\mu} \phi_3
\right)
,
\\
%==================================================
{\cal L}^{(2)}
=&
+
\frac{1}{4} \rho^{a}_{\mu} \rho^{a \mu} 
\left(
g_1^2 s^2 \phi_2^{\dagger} \phi_2
+
g_0^2 c^2 \phi_3^{\dagger} \phi_3
\right)
% \nonumber\\
% %----------------------------------------
%&
+
\frac{1}{8} 
(v_1 + h_1)^2 g_\rho^2
\rho^{a}_{\mu} \rho^{a \mu} 
\nonumber\\
%----------------------------------------
&
-ig
\tr
\left(
[\rho_{\mu}, \rho_{\nu}] W^{\mu \nu}
\right)
\nonumber\\
%----------------------------------------
&- \frac{1}{2} \tr\left(
(D_{\mu}\rho_{\nu} - D_{\nu}\rho_{\mu})
(D^{\mu}\rho^{\nu} - D^{\nu}\rho^{\mu})
\right)
\nonumber\\
%----------------------------------------
&
- i (g_0 c^3 + g_1 s^3) \tr\left(
[\rho_{\mu}, \rho_{\nu}]
(D^{\mu}\rho^{\nu} - D^{\nu}\rho^{\mu})
\right)
\nonumber\\
%----------------------------------------
&+ \frac{1}{2} (g_0^2 c^4 + g_1^2 s^4) \tr\left(
[\rho_{\mu}, \rho_{\nu}]
[\rho^{\mu}, \rho^{\nu}]
\right).
\end{align}
$V(h_1, \phi_2, \phi_3)$ is the scalar potential for the effective lagrangian,
given by substituting the scalar fields to Eq.~(\ref{eq;HiggsPotential}).
Now, $h_1$ is gauge singlet, and $\phi_2$ and $\phi_3$ are SU(2)$_L$-doublet.
Then, we successfully derive a 2HDM with an extra singlet scalar boson and an extra SU(2)$_L$ vector boson.
Note that U(1)$_2$ is identified to U(1)$_Y$ in the SM.

Now, we assume that the breaking scale of SU(2)$_1 \times$SU(2)$_2$ is much higher than 
the EW scale. We find that $\rho^\mu$ and $h_1$ gain the mass proportional to $v_1$ 
so that they become extremely heavy and can be integrated out much above the EW scale under this assumption. Then the effective lagrangian around the EW scale can be described as
%We can neglect higher dimensional operators generated by $\rho^{\mu}$
%and $h_1$.
%After integrating out $\rho$ and $h_1$, we find
\begin{align}
 {\cal L}
=&
{\cal L}^{(0)}
+
V_{eff}(\phi_2, \phi_3)
+
\text{(higher dimensional operators)}
.
\end{align}
Defining the VEVs of $\phi_2$ and $\phi_3$ as $v_2/\sqrt{2}$ and $v_3/\sqrt{2}$,
$V_{eff}(\phi_2, \phi_3)$ is expressed as below:
\begin{align}
V_{eff}(\phi_2, \phi_3)
=&
- \kappa \frac{v_3 v_1}{4 v_2} 
 \left( \phi_2^{\dagger} \phi_2 - \frac{v_2^2}{2} \right) 
- \kappa \frac{v_1 v_2}{4 v_3} 
 \left( \phi_3^{\dagger} \phi_3 - \frac{v_3^2}{2} \right) 
+
\kappa \frac{v_1}{4} 
 (\phi_2^{\dagger} \phi_3 + \phi_3^{\dagger} \phi_2)
\nonumber\\
%------------------------------
&
+ 
\left(
\tilde{\lambda}_2 + \frac{v_1^2 \tilde{\lambda}_{12}^2}{2 m_{h_1}^2}
\right)
\left( \phi_2^{\dagger} \phi_2 - \frac{v_2^2}{2} \right)^2 
+ 
\left(
\tilde{\lambda}_3 + \frac{v_1^2 \tilde{\lambda}_{31}^2}{2 m_{h_1}^2}
\right)
\left( \phi_3^{\dagger} \phi_3 - \frac{v_3^2}{2} \right)^2 
\nonumber\\
%------------------------------
&
+ 
\left(
\tilde{\lambda}_{23} 
+ \frac{v_1^2}{m_{h_1}^2}  \tilde{\lambda}_{12}  \tilde{\lambda}_{31} 
\right)
   \left( \phi_2^{\dagger} \phi_2 - \frac{v_2^2}{2} \right)
   \left( \phi_3^{\dagger} \phi_3 - \frac{v_3^2}{2} \right) 
\nonumber\\
%------------------------------
&
+
\frac{\kappa^2}{32 m_{h_1}^2}
\left(
\phi_2^\dagger \phi_3  + \phi_3^\dagger \phi_2  - v_2 v_3
\right)^2
\nonumber\\
%------------------------------
&
+
\frac{\kappa v_1}{4 m_{h_1}^2} \tilde{\lambda}_{12}
\left( \phi_2^\dagger \phi_2  - \frac{v_2^2}{2} \right)
\left( \phi_2^\dagger \phi_3 + \phi_3^\dagger \phi_2 -v_2 v_3 \right)
\nonumber\\
%------------------------------
&
+
\frac{\kappa v_1}{4 m_{h_1}^2} \tilde{\lambda}_{31}
\left( \phi_3^\dagger \phi_3  - \frac{v_3^2}{2} \right)
\left( \phi_2^\dagger \phi_3 + \phi_3^\dagger \phi_2 -v_2 v_3 \right)
.
\end{align}
We can easily find out the correspondence between the Higgs potential in  Eq.(\ref{eq:2hdm_potential})
and $V_{eff}(\phi_2, \phi_3)$, according to the following identifications,
\begin{align}
&\phi_2 = \Phi_1, \ \ v_2 = v_d, \\
&\phi_3 = \Phi_2, \ \ v_3 = v_u.
\end{align}
The each parameter in $V_{eff}$ corresponds to the one in 
Eq.~(\ref{eq:2hdm_potential}) as follows:
\begin{align}
 m_1^2
=&
 -\frac{\kappa v_1 v_3}{4 v_2}
 - v_2^2 \left(\tilde{\lambda}_2 
   + \frac{v_1^2 \tilde{\lambda}_{12}^2}{2 m_{h_1}^2} \right)
 - \frac{v_3^2}{2} \left(\tilde{\lambda}_{23} 
   + \frac{v_1^2 \tilde{\lambda}_{12} \tilde{\lambda}_{31}}{m_{h_1}^2} \right)
 - \frac{\kappa v_1 v_2 v_3}{4 m_{h_1}^2} \tilde{\lambda}_{12}
,\\
%==============================
 m_2^2
=&
 -\frac{\kappa v_1 v_2}{4 v_3}
 - v_3^2 \left(\tilde{\lambda}_3 
   + \frac{v_1^2 \tilde{\lambda}_{31}^2}{2 m_{h_1}^2} \right)
 - \frac{v_2^2}{2} \left(\tilde{\lambda}_{23} 
   + \frac{v_1^2 \tilde{\lambda}_{12} \tilde{\lambda}_{31}}{m_{h_1}^2} \right)
 - \frac{\kappa v_1 v_2 v_3}{4 m_{h_1}^2} \tilde{\lambda}_{12}
,\\
%==============================
 m_3^2
=&
 \frac{\kappa v_1}{4}
- \frac{\kappa^2 v_2 v_3}{16 m_{h_1}^2} 
- \frac{\kappa v_1 v_2^2}{8 m_{h_1}^2} \tilde{\lambda}_{12}
- \frac{\kappa v_1 v_3^2}{8 m_{h_1}^2} \tilde{\lambda}_{31}
,\\
%==============================
\lambda_1
=&
 2 \tilde{\lambda}_2 + \frac{v_1^2 \tilde{\lambda}_{12}^2}{ m_{h_1}^2}
,\\
%==============================
\lambda_2
=&
 2 \tilde{\lambda}_3 + \frac{v_1^2 \tilde{\lambda}_{31}^2}{ m_{h_1}^2}
,\\
%==============================
\lambda_3
=&
 \tilde{\lambda}_{23} 
 + \frac{v_1^2}{ m_{h_1}^2} \tilde{\lambda}_{12} \tilde{\lambda}_{31}
,\\
%==============================
\lambda_4
=
\lambda_5
=&
 \frac{\kappa^2}{ 16 m_{h_1}^2}
\simeq
0  \label{eq;lambda4}
,\\
%==============================
\lambda_6
=&
 \frac{\kappa v_1}{ 4 m_{h_1}^2} \tilde{\lambda}_{12}
\simeq
0
,\\
%==============================
\lambda_7
=&
 \frac{\kappa v_1}{ 4 m_{h_1}^2} \tilde{\lambda}_{31}
\simeq
0
.
\end{align}
The relation, $\lambda_{4}=\lambda_{5}$, is respected, so that $m^2_{H^{\pm}}=m^2_A$ is satisfied,
according to Eq. (\ref{eq;massdifference}). 
Since $m_{h_1}^2$ is ${\cal O}(v_1^2)$ and much heavier than the other
dimensional parameters, we can conclude that $\lambda_{4}$ and $\lambda_{5}$ are very tiny.
 Similarly, $\lambda_{6}$ and $\lambda_{7}$ are almost vanishing, in our model.
Therefore $V_{eff}$ is the same as the Higgs potential in 2HDM with
softly broken $Z_2$ symmetry, keeping CP symmetry.
Besides, $V_{eff}$ leads the degenerated
masses for the CP-odd and the charged Higgs bosons.
%These three natures are often assumed in analysis of 2HDM. 
Then we conclude that our model naturally explains the origins of the three conditions, (i)-(iii), that are
often assumed in studies of 2HDMs.\footnote{Our model leads $m_{H^{\pm}}
 = m_A^2$. We need other models in order to lead $m_{H^{\pm}} = m_{H^{0}}^2$.
 Phenomenology with a light pseudoscalar in the latter condition is
 discussed, for example, in Ref.~\cite{Cervero:2012cx}.}

\subsection{Emergence of the Inert doublet model}
\label{sec:inert_2HDM}
In this section, we consider one specific scenario, in the framework of the Type-I 2HDM. 
In general, the two Higgs doublets gain non-vanishing VEVs,
but, in fact, $\phi_2$ need not develop a nonzero VEV because $\phi_2$ does not couple with the SM fermions. We focus on the scenario with $v_2 = 0$ below.

This kind of model is called the Inert 2HDM~\cite{Deshpande:1977rw,
Barbieri:2006dq}. 
This model predicts
a stable neutral particle which can be a dark matter candidate. 
This is an attractive feature of this model.
 $v_2 = 0$ is realized by $\kappa = 0$, as we see from
 Eq. (\ref{eq;HiggsPotential}).
 Then, we write down the Higgs potential in this scenario: 
\begin{align}
V(h_1, \phi_2, \phi_3)
=&
\tilde{\lambda}_3 
\left( \phi_3^{\dagger}\phi_3 -  \frac{v_3^2}{2}  \right)^2
\nonumber\\
%------------------------------
&
+ \left(\mu_2^2 + \frac{1}{2} \tilde{\lambda}_{12} v_1^2 \right)
  \phi_2^{\dagger} \phi_2
+ \tilde{\lambda}_2 (\phi_2^{\dagger} \phi_2)^2
+ \tilde{\lambda}_{23} (\phi_2^{\dagger} \phi_2)(\phi_3^{\dagger} \phi_3)
\nonumber\\
%------------------------------
&
- \frac{1}{4} \tilde{\lambda}_{31} v_3^2 h_1^2 
+ \frac{1}{4} \tilde{\lambda}_1 \left( h_1^2 + 2 v_1 h_1 \right)^2
\nonumber\\
%------------------------------
&
+
\frac{1}{2}
( h_1^2 + 2 v_1 h_1 )
\left(
\tilde{\lambda}_{12}
\phi_2^{\dagger} \phi_2
+
\frac{1}{2}
\tilde{\lambda}_{31}
\phi_3^{\dagger} \phi_3
\right)
.
%\label{eq:lag_i2hdm}
\end{align}
Even if $h_1$ is integrated out, $\lambda_4$ and
$\lambda_5$ cannot be induced, as we see in Eq. (\ref{eq;lambda4}).
Thus $\lambda_4 = \lambda_5 =0$ is satisfied at the tree level and
this leads degenerate masses for the extra scalars of $H_2$. 

It is known 
that 
the degenerated CP-even and -odd neutral scalars 
enhance 
the cross section of the direct search for dark matters
via the inelastic scattering of the neutral particles with nucleus
through $Z$-boson exchanging~\cite{Barbieri:2006dq}.
This enhancement is enough large to exclude models. Then we have to
find ways to split the masses of the two neutral scalars.

Naive expectation is that loop corrections split the masses. 
However,
this does not happen due to an accidental U(1) symmetry. 
The original Lagrangian without the $\kappa$ term has an accidental
symmetry under which $H_2$ transforms as $H_2 \to \exp(i \alpha_2) H_2$ and all
the other fields do not transform. This symmetry remains even at the low
energy as a global U(1) symmetry only for $\phi_2$.
This global symmetry allows $\lambda_4$ term, and we can expect the
charged scalar mass is different from the other neutral scalar masses at
loop level. On the other hand, this global symmetry forbids the
$\lambda_5$ term, and thus the masses of the two neutral scalars keep
degenerated even at loop level.
Therefore our Inert 2HDM is excluded by the direct search for dark matters, 
if the relic abundance of dark matter is dominated by the neutral components of $\phi_2$.

\section{Extension of the model}
\label{sec:type-3_2HDM}

In the previous section, we have discussed BSMs with extended gauge
symmetry which induce the type-I and the Inert 2HDMs as low-energy
effective models.
Based on the above discussion, we try to construct a model that leads other
types of the 2HDMs at low energy. 

In other types, not one but two Higgs doublet fields couple with the SM
fermions, so that we cannot easily extend the gauge symmetry, under
which the Higgs doublets are charged. 
We need some modifications of our model with SU(2)$_0 \times
$SU(2)$_1 \times$U(1)$_2$. 
In this section, we discuss some example ways to modify and extend our model.

\subsection{Emergence of the type-III two-Higgs doublet model}

For the type-II, -X, -Y, and -III two-Higgs doublet models, we need more than one
Yukawa interaction terms. Then, not only $\phi_3$ but also $H_1$ and $H_2$ should be involved in the Yukawa interactions with the SM fermions. 
That is achieved if we allow the
dimension-5 operators, $\bar{\psi}_L H_1 H_2 \psi_R$. 
%We restrict
%ourselves to start from renormalizable model. 
This dimension-5 operators
are generated if we add new vector-like fermions charged under
SU(2)$_1$ gauge symmetry as shown in
Table~\ref{tab:new_matters}.\footnote{This setup is similar to the models
discussed in Refs.~\cite{0906.5567, 1604.03578}. In those papers, the
third generation in the quark sector is distinguished from the other
generations.}
%
%
%======================================================================
% table for matter contents
%======================================================================
\begin{table}[t]
\centering
\caption{Quantum numbers of the new fermion fields.}
\label{tab:new_matters}
\begin{tabular}{c|cc|cc}
          & $Q_{L}$ & $Q_{R}$ & $L_L$   & $L_R$ \\
\hline \hline
SU(2)$_0$ & $\bf 1$ & $\bf 1$ & $\bf 1$ & $\bf 1$ \\
SU(2)$_1$ & $\bf 2$ & $\bf 2$ & $\bf 2$ & $\bf 2$ \\
U(1)$_2$ & $\frac{1}{6}$ & $\frac{1}{6}$ & $-\frac{1}{2}$ & $-\frac{1}{2}$ \\
SU(3)$_c$ & $\bf 3$ & $\bf 3$ & $\bf 1$ & $\bf 1$ 
\end{tabular}
\end{table}
%======================================================================
%
%
%
Using these new fermions and the original fermions, the Yukawa
interaction terms are given by
\begin{align}
{\cal L}^{Yukawa}
=&
- \sum_{i,j} \overline{q}_L^{i}  \tilde{\phi_3} y_u^{ij}  u_R^{j}
- \sum_{i,j} \overline{q}_L^{i}  \phi_3 y_d^{ij}  d_R^{j}
- \sum_{i} \overline{\ell}_L^{i} \phi_3 y_e^{i}  e_R^{i}
\nonumber\\
%----------------------------------------
&
-
\sum_{i,j}
\bar{q}_L^{i}
H_1 Y_{Q1}^{ij} Q_{R}^{j}
-
\sum_{i,j}
\bar{Q}_R^{i}
M_Q^{ij} Q_{L}^{j}
- \sum_{i,j} \bar{Q}_L^{i} \tilde{\phi}_2 Y_{u2}^{ij} u_R^{j}
- \sum_{i,j} \bar{Q}_L^{i} \phi_2 Y_{d2}^{ij} d_R^{j}
\nonumber\\
%----------------------------------------
&
-
\sum_{i,j}
\bar{\ell}_L^{i}
H_1 Y_{L 1}^{ij} Q_{R}^{j}
-
\sum_{i,j}
\bar{L}_R^{i}
M_L^{ij} L_{L}^{j}
- \sum_{i,j} \bar{L}_L^{i} \phi_2  Y_{2e}^{ij}   e_R^{j}
\nonumber\\
%----------------------------------------
&
+
(h.c.).\label{eq;type3}
\end{align}
We can take $M_Q$ and $M_L$ as diagonal matrices without lose of
generality by the transformation of the vector-like fermions. For
simplicity, we assume that all components of $M_Q$ and $M_L$ are larger
than the other mass parameters in the Yukawa terms, and integrate out the
vector-like fermions. Then we obtain the effective Yukawa couplings for the SM fermions,
\begin{align}
{\cal L}^{Yukawa}
\simeq&
- \sum_{i,j} \overline{q}_L^{i}  \tilde{\phi_3} y_u^{ij}  u_R^{j}
- \sum_{i,j} \overline{q}_L^{i}  \phi_3 y_d^{ij}  d_R^{j}
- \sum_{i} \overline{\ell}_L^{i} \phi_3 y_e^{i}  e_R^{i}
\nonumber\\
%----------------------------------------
&
-
\sum_{i,j}
\bar{q}_L^{i}
H_1 \tilde{\phi}_2 
\left( Y_{Q1} M_Q^{-1} Y_{2u} \right)^{ij}
 u_R^{j} 
-
\sum_{i,j}
\bar{q}_L^{i}
H_1 \phi_2 
\left( Y_{Q1} M_Q^{-1} Y_{2d} \right)^{ij}
 d_R^{j} 
\nonumber\\
%----------------------------------------
&
-
\sum_{i,j}
\bar{\ell}_L^{i}
H_1 \phi_2 
\left( Y_{L1} M_L^{-1} Y_{2e} \right)^{ij}
 e_R^{j}
\nonumber\\
%----------------------------------------
&
+
(h.c.).
\end{align}
Now we can clearly see that this type of Yukawa couplings is categorized
as the one in the type-III 2HDM with extra singlet scalar at low energy, after integrating out $W'$ and $Z'$ as we have done in Sec.~\ref{sec:model1}.
In the limit that $v_1 \gg v_2$ and $h_1$ is extremely heavy, we lead the condition (iii): $\lambda_4
\sim \lambda_5 \sim \lambda_6 \sim \lambda_7 \sim 0$.

\subsection{Emergence of the type-II, -X, and -Y two Higgs doublet
  models with a discrete symmetry}
The type-II, -X, and -Y 2HDMs are also generated effectively from the
setup in Eq. (\ref{eq;type3}) by controlling the Yukawa couplings.
For example, if $y_{d} = y_e = Y_{2u} = 0$ are realized, the model
behaves as the type-II 2HDM at low energy. A popular way to forbid unwanted
terms is to assign a discrete symmetry.
% If we adapt $Z_2$ symmetry, illustrative charge assignments for the
% three types are shown in Table~\ref{tab:Z2_2HDM}. 
An illustrative $Z_2$ charge assignments for the
three types are shown in Table~\ref{tab:Z2_2HDM}. 
The $Z_2$ symmetry plays a role in controlling the Yukawa couplings of
$\phi_3$, so that it is the same as the one in the ordinary 2HDMs.
 Besides, these assignments forbid $\kappa$ term which is corresponding
 to the soft mass term in the two-Higgs doublet model. However, $\kappa$
 term is required to avoid spontaneous $Z_2$ symmetry breaking, and thus
 this $Z_2$ symmetry must be broken softly. 
%
%
%
%======================================================================
% table for matter contents
%======================================================================
\begin{table}[t]
\centering
\caption{Example of the $Z_2$ charge assignment for various two-Higgs
 doublet model.}
\label{tab:Z2_2HDM}
\begin{tabular}{c|ccc|cccc|ccc}
%fields & type-II & type-X & type-Y \\
$Z_2$ & $H_1$  & $H_2$  & $\phi_3$ 
& $Q_{L,R}$ & $L_{L,R}$ & $q_L$ & $\ell_L$
& $u_R$     & $d_R$     & $e_R$  \\
\hline \hline
type-II & $\pm$ & $\pm$ & $-$ 
& $\pm$ & $\pm$ & $+$ & $+$
& $-$     & $+$ & $+$  
\\
type-X & $\pm$ & $\pm$ & $-$ 
& $\pm$ & $\pm$ & $+$ & $+$
& $-$     & $-$ & $+$  
\\
type-Y & $\pm$ & $\pm$ & $-$ 
& $\pm$ & $\pm$ & $+$ & $+$
& $-$     & $+$ & $-$  
\\
\end{tabular}
\end{table}
%======================================================================

In the type-II, -X, and -Y 2HDMs which are emerged from this model, the CP-odd and
the charged Higgs are automatically degenerated and the Higgs potential
respects CP symmetry. However, since we introduce $Z_2$ symmetry to forbid
unwanted Yukawa interactions, we can not address the origin of $Z_2$
symmetry as we did in the type-I 2HDM shown in Sec.~\ref{sec:model1}.

\subsection{Emergence of the type-II, -X, and -Y two Higgs doublet
  models with a global symmetry}
There is another way to forbid unwanted terms without $Z_2$ symmetry.
Instead of imposing $Z_2$ symmetry, we impose a global U(1) symmetry. 
In order to forbid some Yukawa interactions,
the right-handed fermions, $H_2$, and $\phi_3$ have to be charged under
this global symmetry. However, this means that 
the global symmetry is spontaneously broken by the Higgs VEVs, and
predicts a NG boson whose decay constant is around the electroweak scale. In that
case, the model is similar to the QCD axion model and already
excluded. 
% Therefore we do not consider to the possibility of global
% symmetries here.

To avoid the constraint, we have to extend our model, and introduce new
gauge singlet scalar $S$. The charge assignment is given in 
Table~\ref{tab:U(1)_2HDM}. Here $x_u \neq x_d$ is required to forbid
some Yukawa interaction terms. There are two choices for
$x_S$. When $x_S = -(x_u + x_d)$ is satisfied, we can write down $\phi_3^{\dagger}
H_1 \phi_2$, while $x_S = -(x_u + x_d)/2$ leads $\phi_3^{\dagger}
H_1 \phi_2 S^{*}$. If $x_S$ does not satisfy both, the soft $Z_2$ breaking term is not
emerged at low energy.
%======================================================================
% table for U(1) assignment
%======================================================================
\begin{table}[tb]
\centering
\caption{Examples of U(1) charge assignments to forbid some Yukawa
 interaction terms. Here $x_u \neq x_d$.
We take $x_S$ as $-(x_u + x_d)$ or $-(x_u + x_d)/2$ to lead the soft
 $Z_2$ breaking term at low energy.
}
\label{tab:U(1)_2HDM}
\begin{tabular}{c|cc|cc|cc|ccc|ccc|c}
U(1) & 
$q_L$ & $Q_L$ & $\ell_L$ & $L_L$ & $Q_R$ & $L_R$ & 
$u_R$ & $d_R$ & $e_R$ &
$H_1$ & $\phi_2$ & $\phi_3$ & $S$
\\ \hline \hline
type-II & 
0 & 0 & 0 & 0 & $x_S$ & $x_S$ & 
$x_u$ & $x_d$ & $x_d$ &
$-x_S$ & $-x_d$ & $x_u$ & $x_S$
\\ \hline
type-X & 
0 & 0 & 0 & 0 & $x_S$ & $x_S$ & 
$x_u$ & $x_u$ & $x_d$ &
$-x_S$ & $-x_d$ & $x_u$ & $x_S$
\\ \hline
type-Y & 
0 & 0 & 0 & 0 & $x_S$ & $x_S$ & 
$x_u$ & $x_d$ & $x_u$ &
$-x_S$ & $-x_d$ & $x_u$ & $x_S$
\end{tabular}
\end{table}
%======================================================================
%
%

Since this model is similar to the DFSZ axion model~\cite{Dine:1981rt}, we expect that
the VEV of $S$ is ${\cal O}(10^{11})$~GeV. Then the VEV of $H_1$ should
be also as large as the VEV of $S$ to reproduce the fermion masses. This
naturally leads the decouplings of $\rho^\mu$ and $h_1$. 
In addition, we can solve the strong CP problem in the type-II and -Y
cases if we choose $x_S = -(x_u + x_d)/2$.

\section{Summary}
\label{sec:summary}

The structure of the SM gives some hints to the new physics behind the SM.
One important prediction of the SM is very small flavor and CP
violations, and another is small deviation of the $\rho$ parameter,
which is realized by the custodial symmetry. 
These aspects strongly constrain the extensions of the SM.

The 2HDMs are widely discussed as candidates for BSMs.
In the analysis of the 2HDMs, there are three features 
to realize the above conditions:
(i) softly broken $Z_2$ symmetry, 
(ii) CP invariant Higgs potential, 
and (iii) degenerated mass spectra for the custodial symmetry.
These three features play a crucial role in  
forbidding flavor violating Higgs interactions and avoiding large
contribution to the $\rho$ parameter. Besides, they usually simplify
setups and analyses.
However, their origins are unclear, and they look artificial from a 
viewpoint of bottom-up approach.

In this paper, we have proposed a model with extended electroweak gauge
symmetry, SU(2)$_0 \times$SU(2)$_1 \times$U(1)$_2$, to explain the
origins of the three conditions.  
We have shown that 
the low energy behavior of the model is well described by the
type-I 2HDM with the three features.
We also have discussed the extension of the models to derive other types
of the 2HDM, and shown that the extended models can also solve the strong CP problem, 
imposing a global U(1) symmetry.

\section*{Acknowledgments}
This work was supproted by 
JSPS KAKENHI Grant Number 16K17715.
%
%

%\appendix

\end{document}